\input epsf

    \documentstyle[nato]{crckapb}

\def\figsizeA{2.0}
\def\figsizeB{3.2}

\def\figure#1#2#3{\epsfxsize=#3truein
\vskip -0.2truein
\centerline{\epsffile{fig_#1.eps}}
\centerline{\vbox{{\bf \noindent Figure #1.} #2}}
\medskip}

\def\Psibar{\overline{\Psi}}

\def\PsibarPsi{\langle \overline{\Psi} \Psi \rangle}

\def\meff{m_{\rm eff}}
\def\mres{m_{\rm res}}

\def\MeV{{\rm\  MeV}}

\def\chidof{\chi^2 / {\rm d.o.f.}}
\def\betat{\bar{\beta_1}}

\def\spose#1{\hbox to 0pt{#1\hss}}
\def\ltapprox{\mathrel{\spose{\lower 3pt\hbox{$\mathchar"218$}}
 \raise 2.0pt\hbox{$\mathchar"13C$}}}
\def\gtapprox{\mathrel{\spose{\lower 3pt\hbox{$\mathchar"218$}}
 \raise 2.0pt\hbox{$\mathchar"13E$}}}
\def\inapprox{\mathrel{\spose{\lower 3pt\hbox{$\mathchar"218$}}
 \raise 2.0pt\hbox{$\mathchar"232$}}}

\def\one{
$\PsibarPsi$ in units of the photon mass $m_\gamma$,
vs. $L_s$ for $m_f=0$, fixed physical volume and $m_0 = 0.9$.
From top to bottom the lattice spacing is
$a \sim 1/L$, with $L= 6, 8, 10, 12$.}
\def\two{
$<w> / m_\gamma^2$ vs. $L_s$ for $L=4$ (squares) and 
$L=10$ (diamonds) at $m_0=0.9$.
The physical volume and mass are fixed.}
\def\three{
The large N phase boundary of the parity-flavor broken phase of the 
$SU(2) \times SU(2)$ four-Fermi model on a $6^3$ lattice with $m_0=1$. 
From bottom to top, the $L_s$ values are $2$,
$3$, $4$, and $5$. The parity-flavor symmetry is broken
inside the oval-looking regions.}
\def\four{
$\Psibar \Psi$ vs.\ $m_f$ in the background of a classical instanton
where random noise has been superimposed with 
amplitude $\zeta=0.1$. The lattice size is $16^4$.
The circles, squares, crosses, and diamonds correspond to 
$L_s=4, 6, 8, 10$.}
\def\five{
$<\Psibar \Psi>$ vs.\ $m_f$ from a quenched QCD simulation
above the transition. The lattice size is $16^3 \times 4$,
$\beta= 5.71$, $L_s=32$, and $m_0=1.9$.}
\def\six{
(a) $\PsibarPsi$ vs. $\beta$. (b) $\PsibarPsi$ vs. $m_f$ below and
above the transition.}
\def\seven{
$<\Psibar \Psi>$ vs. $L_s$ below and above the transition.}
\def\eight{
The $\delta$ minus the $\pi$ mass on a $16^4 \times 4$ lattice for
$L_s=16$ and $m_0=1.9$. The fits are to $c_0 + c_2 m_f^2$ and the
stars are the $m_f=0$ extrapolated values.}
\def\nine{
(a) $m_0=1.9$, $m_f=0.02$ and $L_s=24$. The circles correspond
to $16^3 \times 4$ lattice with an ordered initial configuration (ic),
the crosses to a $16^3 \times 4$ with disordered ic,
the diamond to an $8^3 \times 4$ with ordered ic, and
the plus to an $8^3 \times 4$ with disordered ic.
The gauge part of the action
is a Wilson plaquette action. (b) Same as in (a) but with
an Iwasaki improved gauge action with $c_1 = -0.331$.}
\def\ten{
Three flavor QCD with Wilson plaquette action,
$16^3 \times 4$, $m_0=1.9$, $m_f=0.02$ and $L_s=32$. The circles
correspond to an ordered initial configuration and the crosses to a
disordered one.}
\def\eleven{
$m_\pi^2$ extrapolated to $m_f=0$ vs. $L_s$. The circles are for the
Wilson plaquette action and with $m_0=1.9$. The crosses are for the
Iwasaki action with $m_0=1.65$ and the squares with $m_0=1.9$}
\def\twelve{
Histogram of $1 - Tr[U_P]/3$. The area under the curve is normalized
to one. The lattice size is $8^3 \times 4$, $L_s=24$, $m_f=0.02$ and
$m_0=1.9$. (a) Wilson plaquette action at $\beta=5.275$.  (b)
restricted plaquette action with $c=0.8$ at $\beta=1.2$.  Both are
just below the transition.}
\begin{opening}
\title{DOMAIN WALL FERMIONS IN VECTOR THEORIES}

\author{PAVLOS M. VRANAS}
\institute{Physics Department\\
University of Illinois\\
Urbana, IL 61801, USA\\
vranas@uiuc.edu
}

\end{opening}

\begin{document}

\begin{abstract}
Applications of Domain Wall fermions to various vector-like lattice
theories are reviewed with an emphasis on QCD thermodynamics. Methods
for improving their chiral properties at strong
coupling are discussed and results from implementing them are
presented.
\end{abstract}


\section{Introduction}

Domain Wall fermions (DWF) provide an alternative to traditional lattice
fermion methods. Since their introduction to lattice field theory
\cite{Kaplan0,Kaplan1} they have produced a wealth of theoretical and numerical
advances for vector and chiral lattice theories (see \cite{DWF_reviews}
and references therein; also see these proceedings 
\cite{Neuberger_Dubna,Shamir_Dubna,DWF_Dubna}).

DWF are defined in a space-time with an extra dimension.
A mass defect along the extra direction causes one
chirality of the Dirac spinor to get exponentially localized
along the defect. If the extra direction is finite 
an anti-defect will necessarily be present and the other
chirality will also get exponentially localized. As a result,
the two chiralities of a Dirac spinor have been separated
and their mixing is only exponentially small. This 
is suitable to vector-like theories.  The effects
of the defect, anti-defect can also be produced if the
theory is defined without them, but the boundary conditions
along the extra direction are free \cite{Shamir,Furman_Shamir}. 
Then each chirality gets localized along a different boundary (wall).
If the number of sites along the extra direction ``s''
is $L_s$ then the mixing of the two chiralities is
exponentially small in $L_s$. At $L_s = \infty$ the theory
has exact chiral symmetry.

These properties make DWF a powerful tool for numerical studies that
require very good control over the chiral properties.  Consider
dynamical QCD thermodynamics as an example.  At finite lattice spacing
``a'' traditional fermions (staggered or Wilson) break the chiral
symmetry explicitly. The symmetry is recovered together with the
Lorentz symmetry as the lattice spacing tends to zero. But the
computational cost for reducing the lattice spacing is very
large. Typically, in order to reduce the lattice spacing by a factor
of two, one has to perform $2^{8 - 10}$ more computations (the exponent
accounts for the $2^4$ more lattice sites and the various algorithmic
costs).  On the other hand, DWF can approach the chiral limit even at
finite lattice spacing by simply increasing $L_s$. For the first time
the chiral and Lorentz symmetry limits have been separated!
Furthermore the computational cost for increasing $L_s$ is only linear
in $L_s$ making this method very appealing.  Now one has control in
deciding which limit to get closer to.  If the interest is in the
order and properties of the finite temperature phase transition one
would want to bring the theory as close as possible to the chiral
limit and be less concerned with the amount of Lorentz symmetry
present. Of course, other applications may have different requirements.

In this work the action and numerical methods are as in \cite{Furman_Shamir} 
with the modification as in \cite{PMV}. The only notation the
reader will need, to follow the results presented here, is:
$L_s$ is the number of sites along the extra direction ``s'',
$m_0$ is the five dimensional defect mass or domain wall ``height'',
$m_f$ is an explicit mass that mixes the plus chiral component
localized on one wall with the minus component localized in the other wall.
If $L_s = \infty$ the bare fermion mass is proportional to $m_f$.
To put these parameters into perspective, recall that in 
free theory at finite $L_s$ the effective mass is \cite{PMV}:
\begin{equation}
\meff = m_0 (2 - m_0) \left[ m_f + (1-m_0)^{L_s}\right], \ \ \ \ 0 < m_0 < 2 
\label{meff_free}
\end{equation}

The exponentially small ``residual'' mass $\mres = m_0 (2 - m_0)
(1-m_0)^{L_s}$ is a finite $L_s$ effect reflecting the residual
mixing of the chiral components. The range of $m_0$ determines the
number of flavors.  For $m_0 < 0$ zero flavors, $0< m_0 < 2$ one, $2<
m_0 < 4$ four, $4< m_0 < 6$ six, $6< m_0 <8$ four, $8< m_0 < 10$ one,
and $10< m_0$ zero flavors.

A very important aspect of DWF has to do with their exceptional zero
mode properties in the background of topologically non-trivial gauge
fields. Sadly, this is also the reason for many of the difficulties
that arise in their use. Gauge fields are introduced in the standard
fashion in ordinary space-time. They do not have an extra component
and they are the same across the extra direction \cite{Kaplan1,NN0}.
From this point of view, the extra direction can be thought of as an
internal flavor space. This is the basis of the overlap formalism
\cite{NN0,NN1}. One can see that the transfer matrix $T$ along the
extra direction is the same at every ``s'' slice. As a result, the
effect along the extra direction is simply $T^{L_s}$, which at
$L_s=\infty$ is a projection operator on the ground state of the
corresponding Hamiltonian $H$. Therefore, the fermion determinant in
the background of the gauge field will be the overlap of this ground
state with the state at the boundary. The state at the boundary has a
fixed ``filling'' independent of the gauge field background.  In the
presence of a topologically trivial gauge field background the ground
state of $H$ has the same ``filling'' and the overlap of the two
states is not zero. However, the filling level of the ground state of
$H$ can change depending on the background field. If this happens the
overlap with the boundary state is exactly zero indicating the
presence of exact zero modes. The difference of the two filling levels
is an index for this operator. For smooth topologically non-trivial
backgrounds it has been shown that the index theorem is satisfied
\cite{NN1}. Now one can see how this exceptional property is also the
source of problems at finite $L_s$. If a background gauge field is
slowly changed from one sector to another the number of negative
eigenvalues of $H$ will have to change by one. Therefore, one of the
eigenvalues will have to cross zero. When this happens, the transfer
matrix will have eigenvalue equal to one and there will be no decay
along the extra direction and therefore no localization.  The measure
of such ``topology changing'' configurations is zero
\cite{NN1,Furman_Shamir} so the $L_s = \infty$ limit is well
defined. However, configurations in their vicinity will cause very
slow decay rates \cite{EHN_flow}. The effect of these configurations
in numerical simulations is problematic. In order to achieve desired
levels of chiral symmetry, large $L_s$ may be needed depending on the
value of the coupling constant.

In this work several applications of DWF to vector theories are
presented.  Also, some attempts to improve the chiral properties of
DWF so that smaller $L_s$ values can be used in numerical simulations
are described.  The collection of results is not intended as a
summary of the whole field and only reflects the involvement of the
author in the subject.

\section{Schwinger model}

In this section the two flavor Schwinger model is studied using DWF.
This work has appeared in \cite{PMV}.

Before DWF could be used for QCD simulations some basic questions
regarding their properties had to be investigated: (1) In the
interacting theory, is the approach to the chiral limit exponential in
$L_s$?  (2) How does the exponential ``decay'' rate depend on the
lattice spacing?  (3) How large should $L_s$ be?  (4) Are anomalous
effects reproduced?  

The two flavor Schwinger model provides a good
testing ground for these questions.  It is an interacting gauge theory
in which the $U_A(1)$ symmetry is anomalously broken.  The
corresponding t' Hooft vertex $w = [\Psibar_R \Psi_L]^2 + [\Psibar_L
\Psi_R]^2$ is expected to acquire a non-zero VEV at the mass-less
limit. Furthermore, since there is no spontaneous chiral symmetry
breaking, the chiral condensate can serve as a direct probe of chiral
symmetry breaking due to the DWF regulator.

{\epsfxsize=\figsizeB truein
\centerline{\epsffile{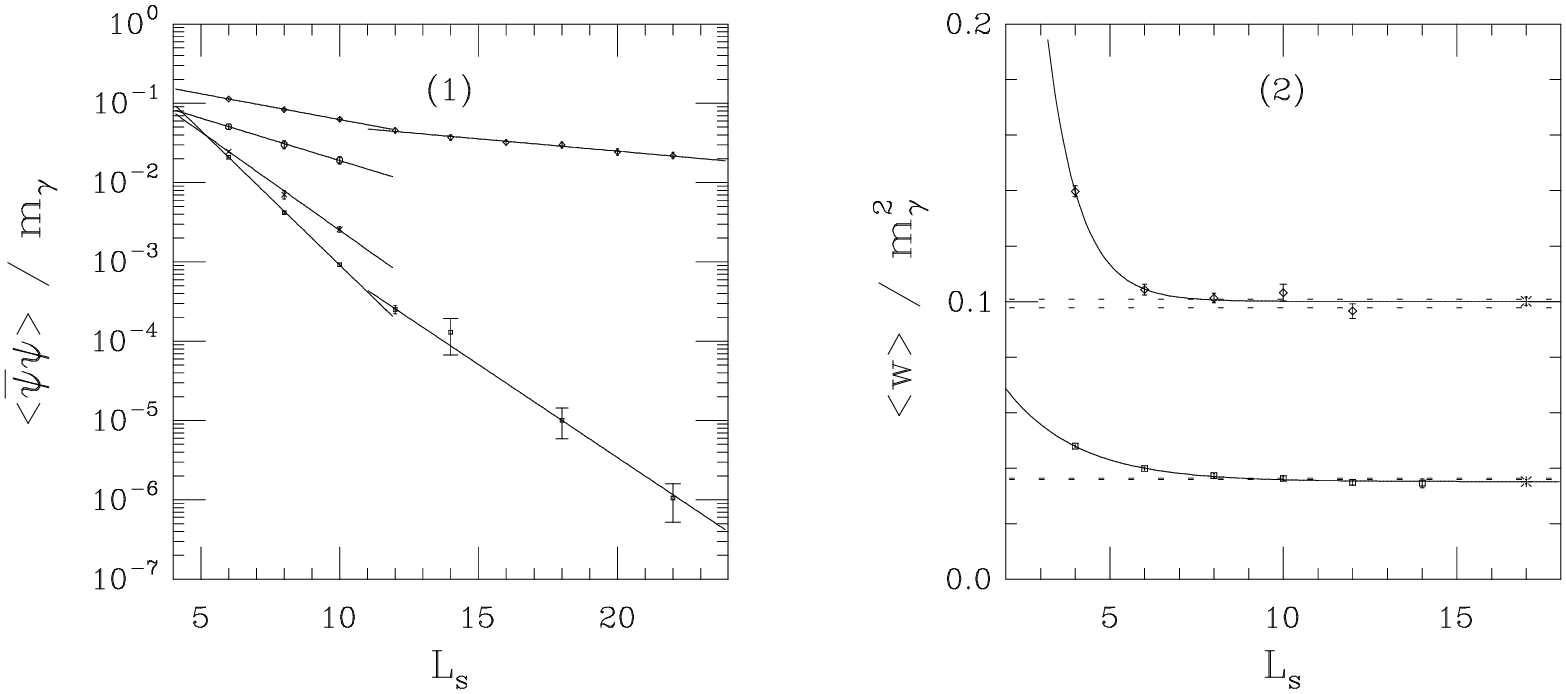}}
\vskip -1.5 truein 
\centerline{\vbox{{\bf \noindent Figure 1.} \one
\hfill\break
{\bf Figure 2.} \two}
\medskip}

The amount of breaking as a function of $L_s$ and the lattice spacing
``a'' as ``seen'' by the chiral condensate $\PsibarPsi$ is shown in
figure 1 for zero $m_f$.  One can see that $\PsibarPsi$ tends to zero
exponentially fast. There are two rates of decay with an inflection
around $L_s=10$. These two rates have been identified in \cite{PMV}
with two separate but related mechanisms. The fast decay can be
thought of as being the free theory decay in the presence of small
fluctuations. For example, one can imagine that in the presence of
small fluctuations $m_0$ in eq. \ref{meff_free} will fluctuate and
therefore even if $m_0=1$ the decay rate will be finite. The slower
decay rate was identified with topology changing configurations of the
type described in the introduction.  It must be pointed out that the
$L_s$ dependence of $\PsibarPsi$ involves many different exponential
decay rates. The two decay rates in figure 1 may in fact be a
superposition of several exponentials. However, it is apparent that
there are two different mechanisms at play and that within the error
bars are nicely summarized by only two exponentials.

The next crucial observation is that the decay rates become faster as
the lattice spacing is reduced. This is very important. If this was
not so then numerical applications of DWF would not be appealing.

In practical applications of DWF one works at non zero $m_f$.  If in a
given application extrapolations make sense then for each value of
$m_f$ one measures at several values of $L_s$ and extrapolates to
$L_s=\infty$ using a function of the form $A + B e^{-c L_s}$.  As an
example, consider a measurement of $w$.  In figure 2, $w$ is measured
vs. $L_s$ for a non-zero mass and fitted to the
above form. The extrapolated value agrees well with the value
obtained using the overlap formalism ($L_s=\infty$ limit of DWF, dotted lines).
Since the overlap has been found to reproduce a non zero VEV for $w$
at $m_f=0$ \cite{NNV,PMV}, one can conclude that DWF
correctly reproduces anomalous effects.

\section{Fermion scalar interactions}

In this section the interaction of DWF with scalar particles is
discussed. This work has appeared in \cite{VKT}.

As discussed in the introduction, the interaction of DWF with gauge fields
was treated by viewing the extra direction as an internal flavor space.
Naively one would think that the interaction of DWF with a spin zero field
could also be treated in the same way. This would imply an interaction of the
form $\sum_s \Psibar(x,s) \sigma(x) \Psi(x, s)$. One can 
see that $\sigma$ will play a role similar to $m_0$. Since the value of
$m_0$ controls the number of flavors in the theory this is clearly the wrong
way to couple the spin zero field. Instead, it should be coupled to the light
degrees of freedom that ``reside'' on the boundaries very much like the
mass term $m_f$: 
\begin{equation}
\Psibar_R(x,0) \sigma(x) \Psi_L(x, L_s-1) + 
\Psibar_L(x,L_s-1) \sigma(x) \Psi_R(x, 0)
\label{dwf_scalar}
\end{equation}
Therefore, it is not natural to think of the extra direction
as an internal flavor space. This reveals some richness in the way
different spin particles interact with DWF. In fact, the outline
of some more ``geometrical picture'' seems to emerge. 

\figure{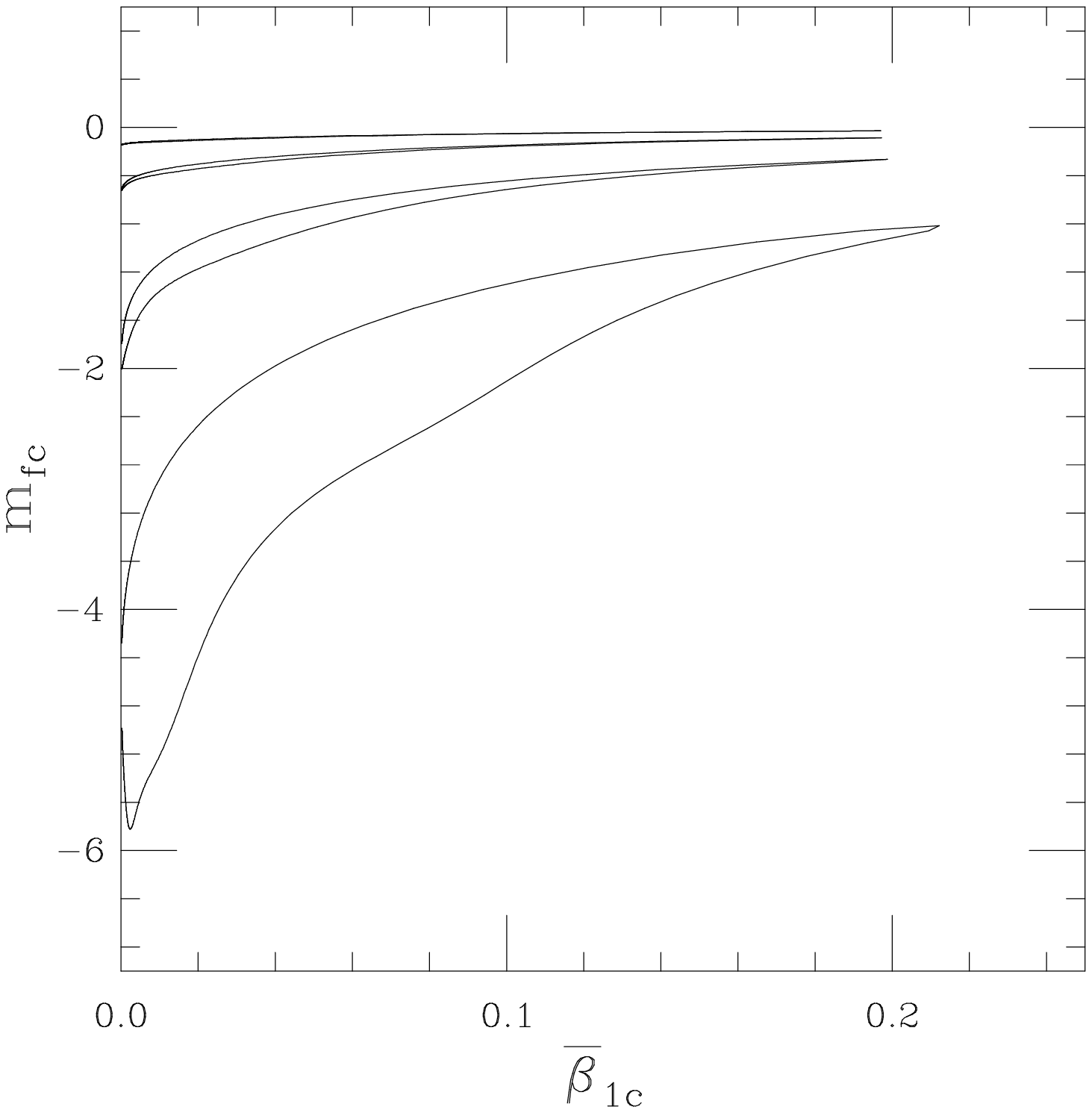}{\three}{\figsizeA}

Using this type of interaction, four-Fermi and Yukawa type
models can be studied with DWF. A simple $Z_2 \times Z_2$
model was studied using large N and numerical techniques.
An interesting result from this study was that the minimum
decay rate of these models as predicted from large N
is $-ln(2 - \sqrt 2) = 0.535$. This was also confirmed
numerically. A similar result has been obtained in \cite{Shamir_Dubna}.

A large N study of the $SU(2) \times SU(2)$ four-Fermi model with DWF
revealed a rich small $L_s$ structure. For small $L_s$ and negative
$m_f$ a parity--flavor broken phase exists. This phase is of the same
nature as the Aoki \cite{Aoki_phase} phase of Wilson fermions.  This
phase was later observed in quenched QCD with DWF \cite{Aoki_phase_DWF}.
The large N phase boundary is shown in figure 3 (where $\betat$ is
proportional to the inverse of the 4-fermi coupling).

\section{Zero modes for classical topological backgrounds}

In this section the zero mode properties of DWF are studied in the
presence of classical instanton backgrounds. This work has appeared in
\cite{CU_zero_modes}.

In the introduction it was described how at $m_f=0$ and at the
$L_s=\infty$ limit DWF can have exact and robust zero modes. This property is
strictly true only at that limit. In practice, one would like to work
at small but finite $m_f$.  It is therefore natural to ask if the zero
mode effects at the small $m_f$ of interest can be reproduced for
reasonable values of $L_s$ and if these effects are stable under
perturbations. 

\figure{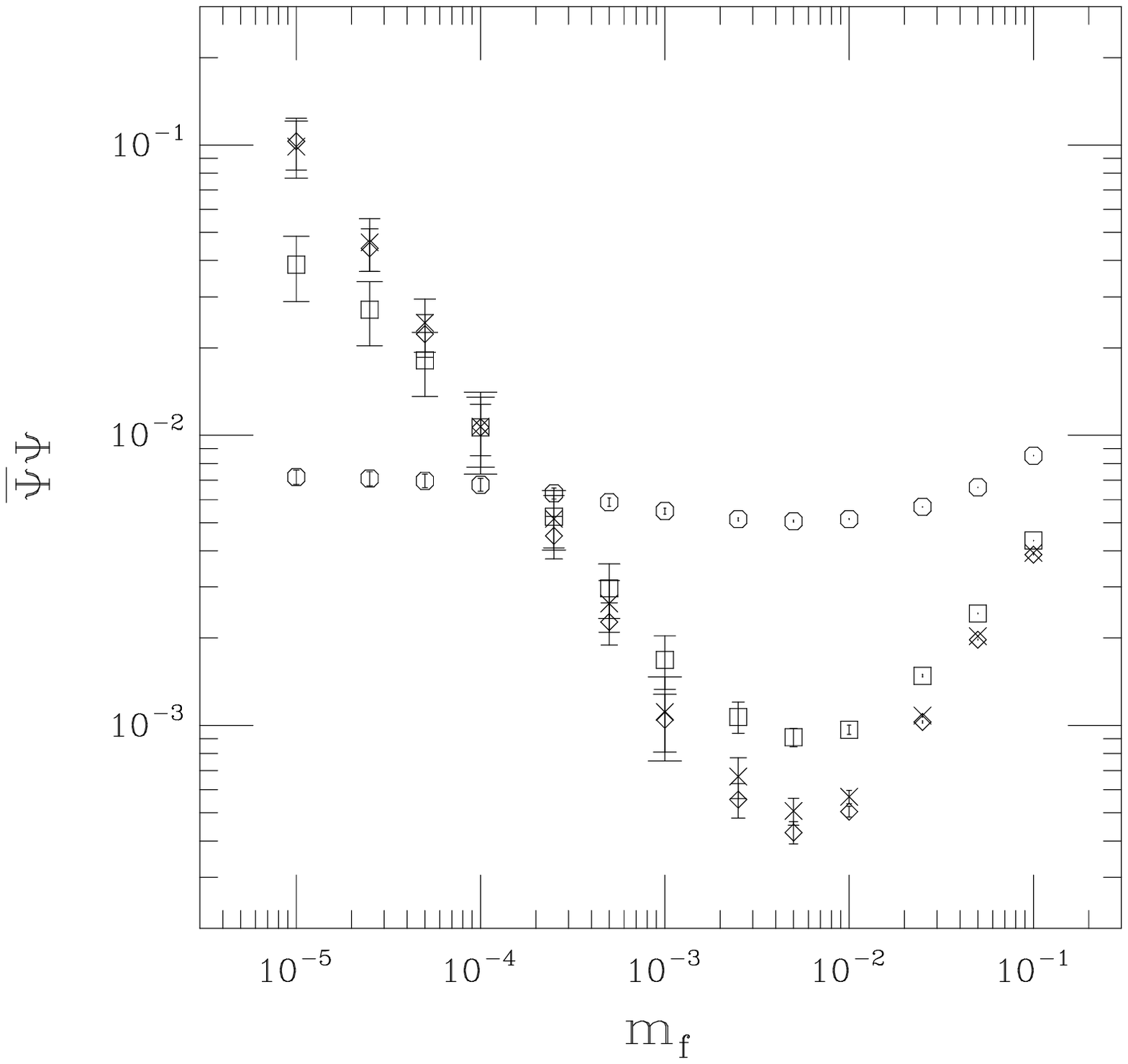}{\four}{\figsizeA}

This question can be addressed by studying the DWF chiral condensate
$\Psibar \Psi$ as a function of $m_f$ in the presence of a background
classical instanton field with some amount of superimposed random
noise with amplitude $\zeta$. If the zero mode effects are present and
are robust under fluctuations then the chiral condensate should
diverge as $1/m_f$. Indeed, this is the case as can be seen from figure
4 where $\zeta=0.1$. For $L_s=4$ the divergence is not visible, for
$L_s=6$ it continues down to $m_f \approx 10^{-4}$ and then it
``flattens out'' indicating finite $L_s$ effects. For $L_s=8, 10$ the
divergence continues down to $m_f \approx 10^{-5}$. Typical
applications do not require $m_f$ below $10^{-3}$.  This result is in
sharp contrast with Wilson or staggered fermions that do not have
robust zero modes.

\section{Quenched QCD above the finite temperature transition}

In this section the effects of zero modes are studied for backgrounds
that involve quantum-noise, namely backgrounds generated
using the pure gauge action above the transition (quenched QCD). 
This has appeared in \cite{lat98_Fleming,lat98_Kaehler}

In a quenched QCD simulation the zero modes are not suppressed by the
fermion determinant. If the simulation is done above the finite
temperature phase transition then one would expect that the chiral
condensate at small masses will not vanish but instead it will diverge
as $1 / m_f$. This would then be a direct indication of the failure of
the quenched approximation. Even so, one would expect that if an
extrapolation is done from large enough masses it will extrapolate to
zero at $m_f=0$ indicating that the theory is in the high temperature
phase.

\figure{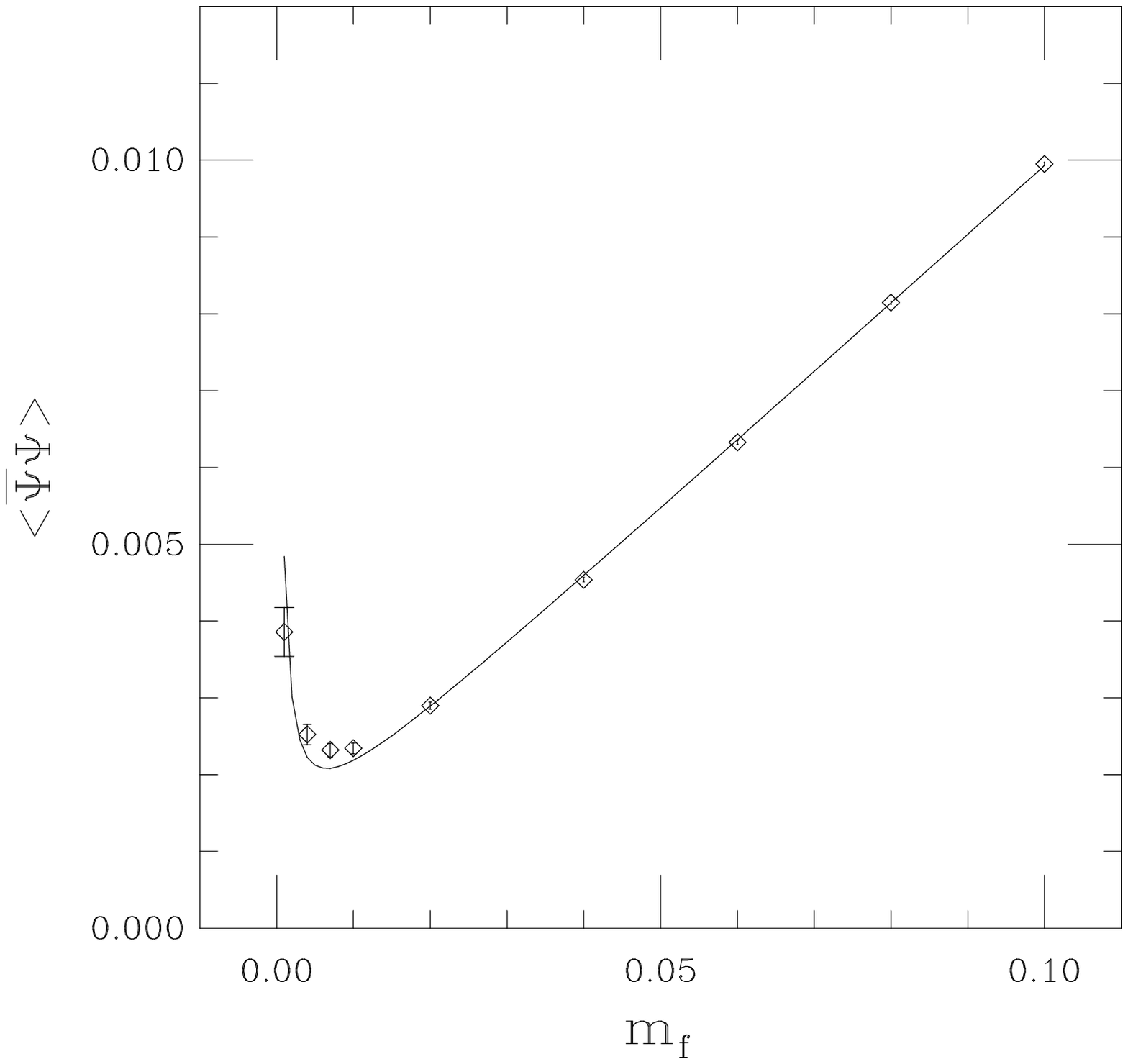}{\five}{\figsizeA}

In figure 5 the chiral condensate as a function of $m_f$ is shown. The
$1 / m_f$ divergence for $m_f < 0.005$ is evident. The fit is to a
form $ c_{-1} / m_f + c_0 + c_1 m_f$.  What is more surprising is that
$c_0$ is not zero but $c_0 = 9.0(4) 10^{-4}$.  This is the first time
such a phenomenon has been observed and is another indication of the
failure of the quenched approximation (recently this was reproduced
using different methods \cite{EHKN_quenched_thermo,lat99_Sinc}). This
indicates that the remaining instantons, anti-instantons, in the
absence of a fermion determinant, are capable in producing a non-zero
density of small eigenvalues.  Again, this is in sharp contrast with
Wilson or staggered fermions.

\section{QCD thermodynamics with DWF}

In this section the parameter space of DWF for the
two flavor QCD finite temperature transition
is studied. This has appeared in 
\cite{lat98_Vranas,ICHEP_NHC,dpf99_Vranas,PANIC99_Fleming}.

One would like to verify that DWF can be used in thermodynamic studies
and to determine the range of the new parameters $m_0$ and $L_s$. A
study of the full two flavor QCD on small lattices $8^3 \times 4$ is
done in order to answer these questions.

{\epsfxsize=\figsizeB truein
\centerline{\epsffile{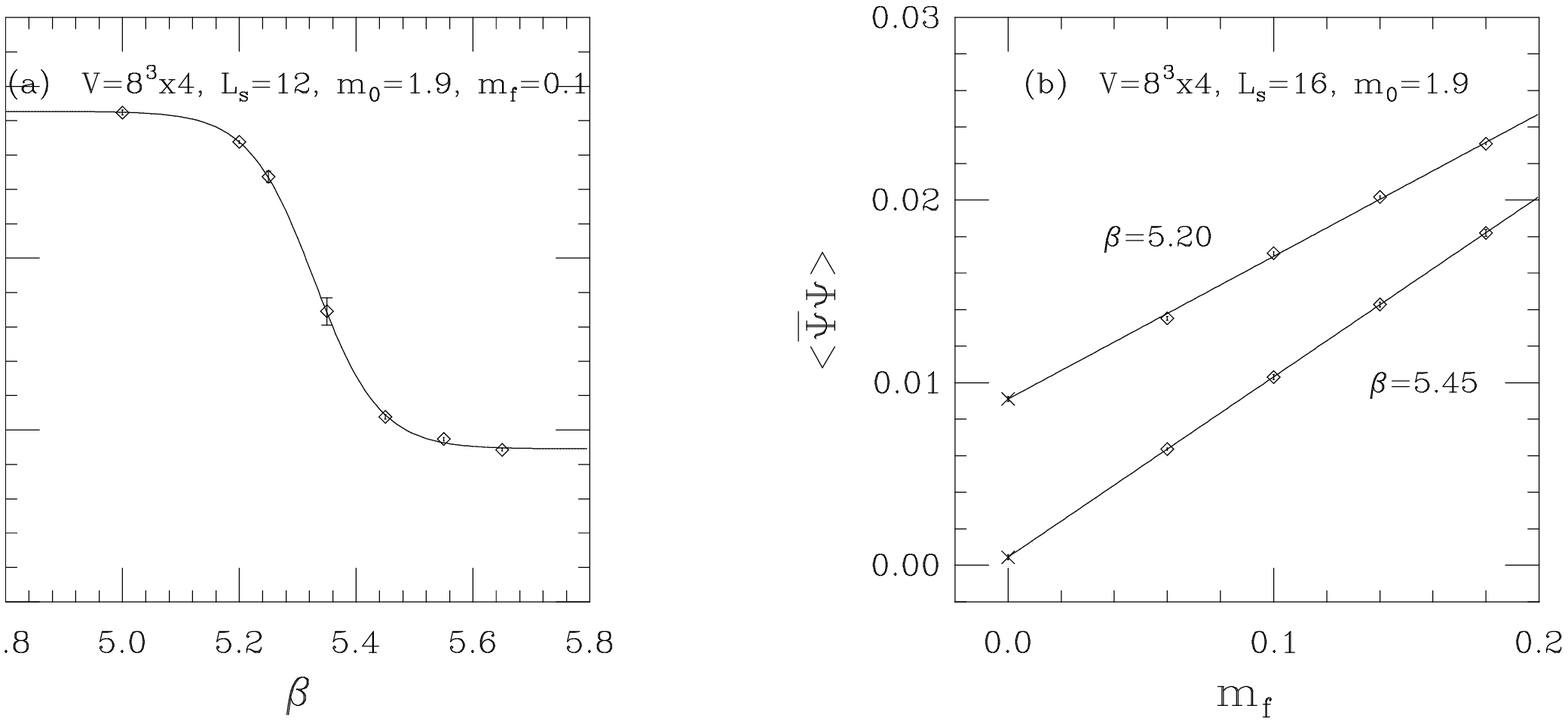}}
\vskip -1.5 truein 
\centerline{\vbox{{\bf \noindent Figure 6.} \six}}
\medskip}

\figure{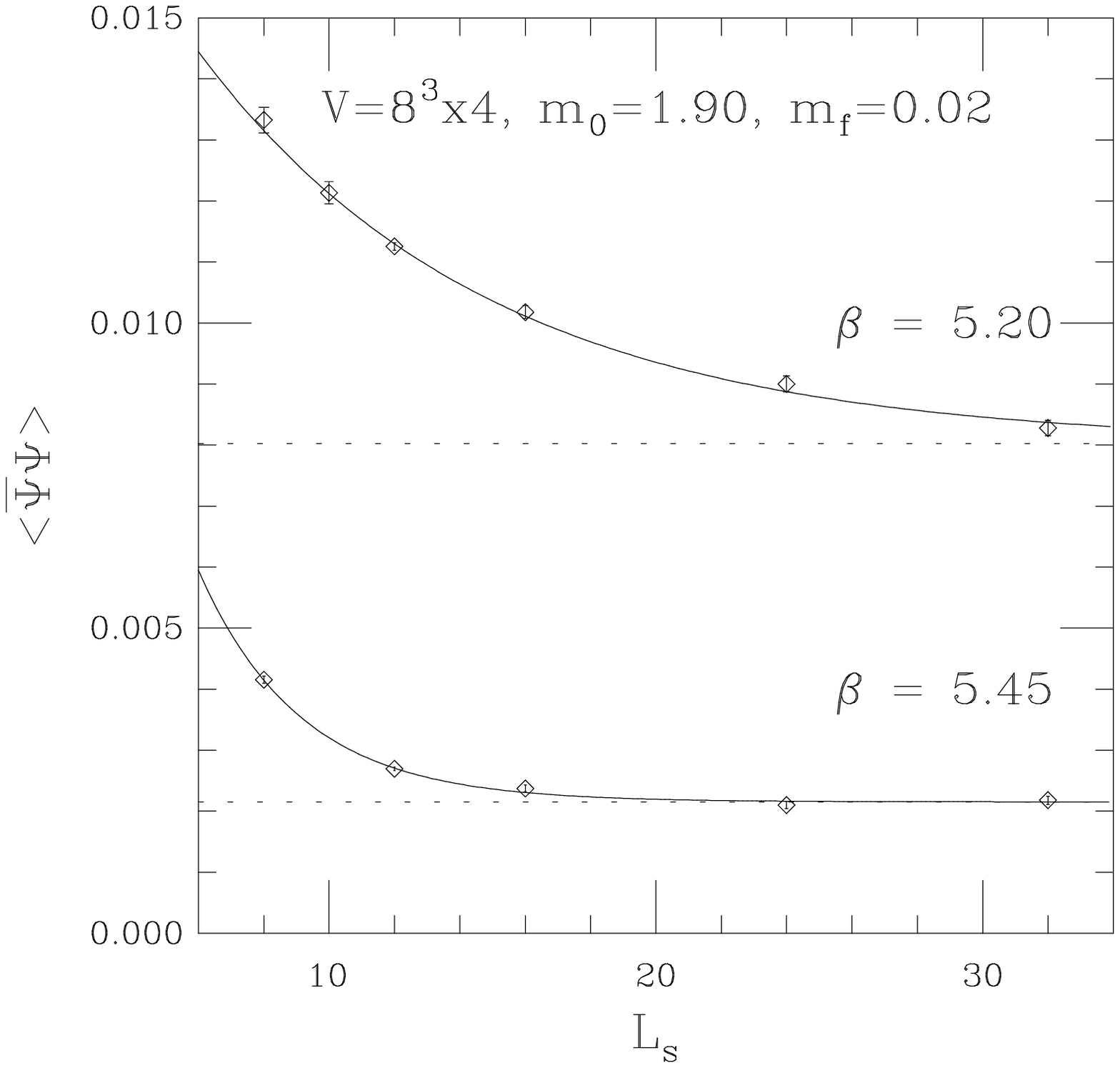}{\seven}{\figsizeA}

In figure 6a, for $m_f=0.1$ and $L_s = 12$, one can clearly see a sharp
crossover in the chiral condensate as $\beta$ is varied.  In figure 6b
$\PsibarPsi$ is plotted vs. $m_f$ for $L_s=16$ at $\beta=5.2$ below the
transition and at $\beta=5.45$ above the transition. Below the
transition it extrapolates to a non zero value and above to a near
zero value. In figure 7 $\PsibarPsi$ is plotted vs. $L_s$ at
$m_f=0.02$.  As can be seen, the decay is consistent to exponential. At
$L_s=24$ $\PsibarPsi$ has reached its asymptotic value above the
transition while below the transition it is $\approx 10 \%$ away.  The
dependence on $m_0$ has also been investigated
\cite{lat98_Vranas,dpf99_Vranas} and has been found that if $m_0$ is
constrained in the region $1.6 < m_0 < 2.0$ then the theory has one
flavor (in our case two, since we simulate with the square of the
determinant of the Dirac operator).  Below that region the light
states disappear and above more flavors may be present.

Since DWF have explicit flavor symmetry, one can see that above the
transition the full $SU(2) \times SU(2)$ chiral symmetry is
present to a very good approximation at $L_s=24$.  Below the
transition the symmetry is spontaneously broken. For the first time
the important symmetry properties of QCD at finite temperature are
under control.

\section{$U_A(1)$ Above the QCD finite temperature phase transition}

The results from a study of the $U_A(1)$ symmetry just above 
the deconfinement transition are presented. This has appeared in 
\cite{lat98_Vranas,ICHEP_NHC,dpf99_Vranas,lat99_Vranas,PANIC99_Fleming}.

As is well known the $U_A(1)$ axial symmetry of zero temperature QCD
is broken by the anomaly. However, at finite temperature above the
transition where the chiral symmetry is restored $U_A(1)$ may also be
restored. This is an interesting subject because there are no
experimental results relating to this yet and because the order of the
transition may depend on it. Lattice simulations could contribute
something important in this direction.  Unfortunately, staggered
fermions \cite{CU_omega,Kogut_Lagae_Sinc} could not produce conclusive
results because they do not have robust zero modes
\cite{Kogut_Lagae_Sinc}.

\figure{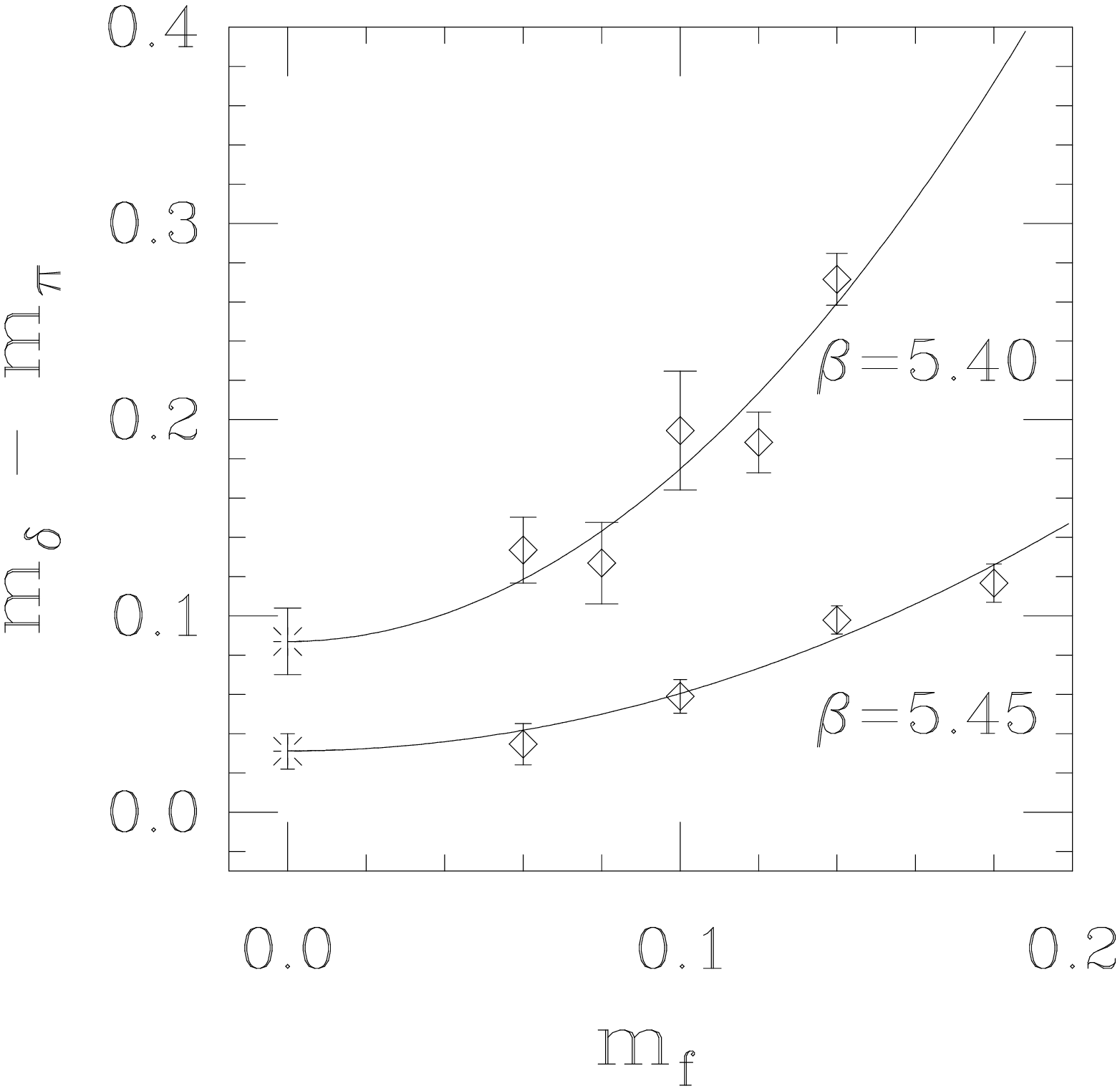}{\eight}{\figsizeA}

On the other hand, DWF have exceptionally good zero mode properties as
discussed in the previous sections. Therefore, they provide a powerful
tool for such studies. The results from a large scale simulation of
the full two flavor QCD on large volumes is presented in figure 8.
The difference of the screening masses of the $\delta$ and $\pi$ is
used as a measure of anomalous symmetry breaking and is plotted
vs. the bare quark mass $m_f$ on a $16^3 \times 4$ lattice for $L_s=16$
at $\beta=5.45$ and $\beta=5.40$ ($\beta_c \approx 5.325$).  The
lines are fits to $c_0 + c_2 m_f^2$ and have $\chidof \approx 1$.  The
fact that a linear term is not needed in order to give a good fit is
an indication that for $L_s=16$ the chiral symmetry is effectively
restored (this can also be seen from figure 7).  The $m_f=0$
extrapolated values are $0.087(17)$ at $\beta=5.40$ and $0.031(9)$ at
$\beta=5.45$.  Although both are not zero by a statistically
significant amount their value is small when compared with $m_\delta$
and $m_\pi$ which are $\approx 1.3$.  Universality arguments require
that if the QCD phase transition is to be second order, the anomalous
$U(1)_A$ must be broken.  It is an open question as to whether the
small size of the $U(1)_A$ symmetry breaking seen here is sufficient
to support this theoretical prediction that the two-flavor QCD phase
transition is second order \cite{Pizarski}.

\section{The transition region of two flavor QCD}
In this section a study of the transition region of two flavor full QCD is
presented. This work has appeared in 
\cite{dpf99_Vranas,lat99_Vranas,lat99_Wu,lat99_RDM}

{\epsfxsize=\figsizeB truein
\centerline{\epsffile{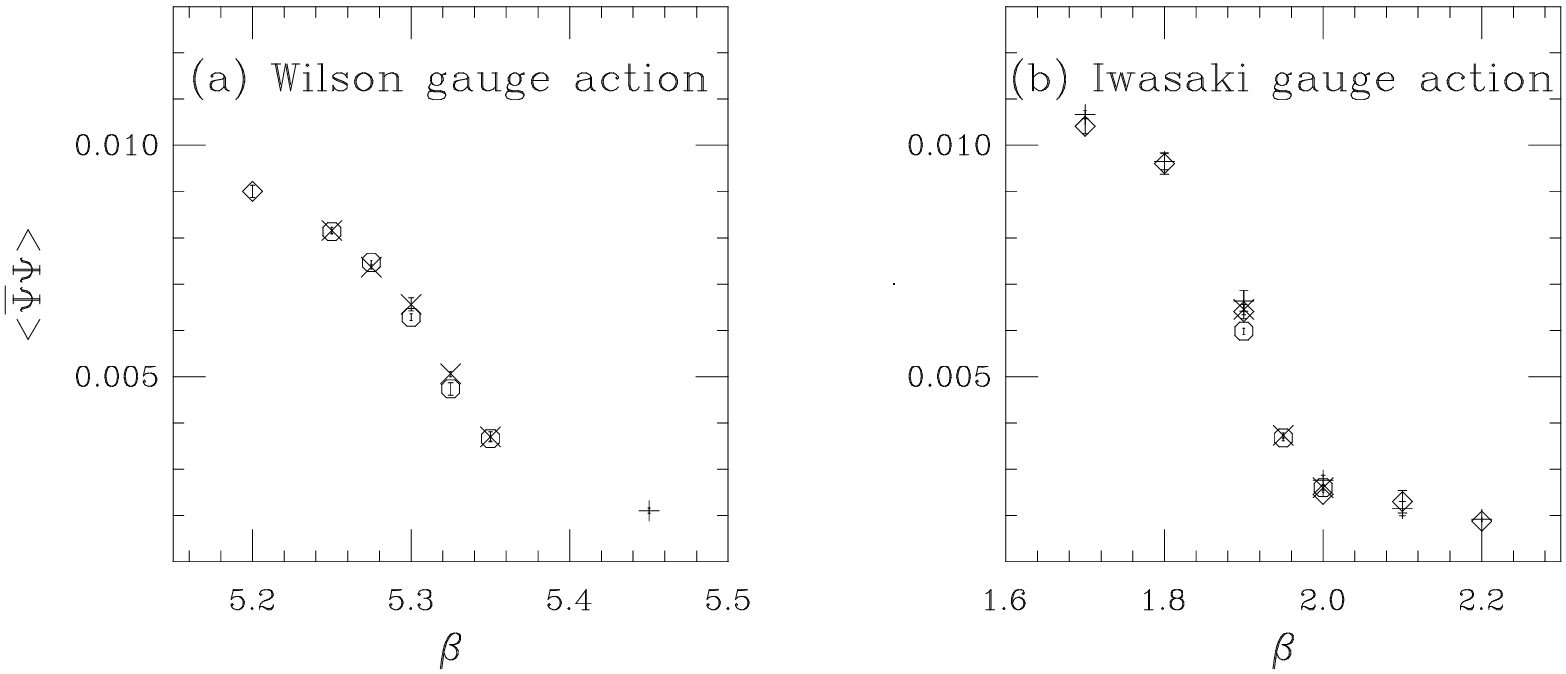}}
\vskip -1.5 truein 
\centerline{\vbox{{\bf \noindent Figure 9.} \nine}}
\medskip}

From the analysis in the previous sections it appears that DWF can
reproduce the important physics of the deconfining transition. Also,
the previous analysis indicates that $L_s=24$, $m_f=0.02$ and $m_0=1.9$ are
reasonable settings for the new parameters.  Armed with this knowledge
a large scale numerical study of the transition region was done and
the results are shown in figure 9a.  The lattice size is $16^3 \times
4$ which corresponds to a physically relevant volume. The transition
appears to be relatively smooth. In order to set the scale a zero temperature
calculation was done at the transition coupling on an $8^3\times 32$
lattice \cite{lat99_Wu}. The scale was set using the $\rho$ mass. 
The critical temperature
was found to be $T_c = 163(4) \MeV$ and the pion mass $m_\pi = 427(11) \MeV$.
The critical temperature is in agreement with results obtained from
other fermion regulators \cite{lat99_Karsch}. The pion mass is clearly
too heavy. More sophisticated studies
\cite{lat98_Fleming,lat99_Fleming} indicate that the residual chiral symmetry
breaking effects are much larger than expected and in order to obtain
a physical pion mass $L_s \approx 64 - 96$ may be needed.

On the other hand, one would expect that the use of a pure
gauge action with better continuum limit properties may improve the
chiral properties of DWF for the same $L_s$.  The previous results
were obtained using the traditional Wilson plaquette action. A
second large scale simulation was done using the Iwasaki improved pure
gauge action with $c_1=-0.331$ \cite{Iwasaki}. The results are shown
in figure 9b.  A similar scale setting calculation \cite{lat99_Wu} 
at $\beta=1.9$ gave
$T_c = 166(3) \MeV$ and $m_\pi = 400(7) \MeV$.  The critical
temperature is in agreement with the Wilson plaquette action but the
pion mass did not get lighter by a significant amount.

The transition at these parameter values is clearly not first order.
Simulations using ordered and disordered initial configurations
at the transition point agreed after a few hundred iterations and 
did not show any signs of first order behavior. Therefore, one can
conclude that the order of the two flavor QCD transition
with three pions of mass $\approx 400$MeV is not first order.

\section{The transition region of three flavor QCD}

Here the three flavor transition is discussed.
This work is in progress \cite{CU_group}.

\figure{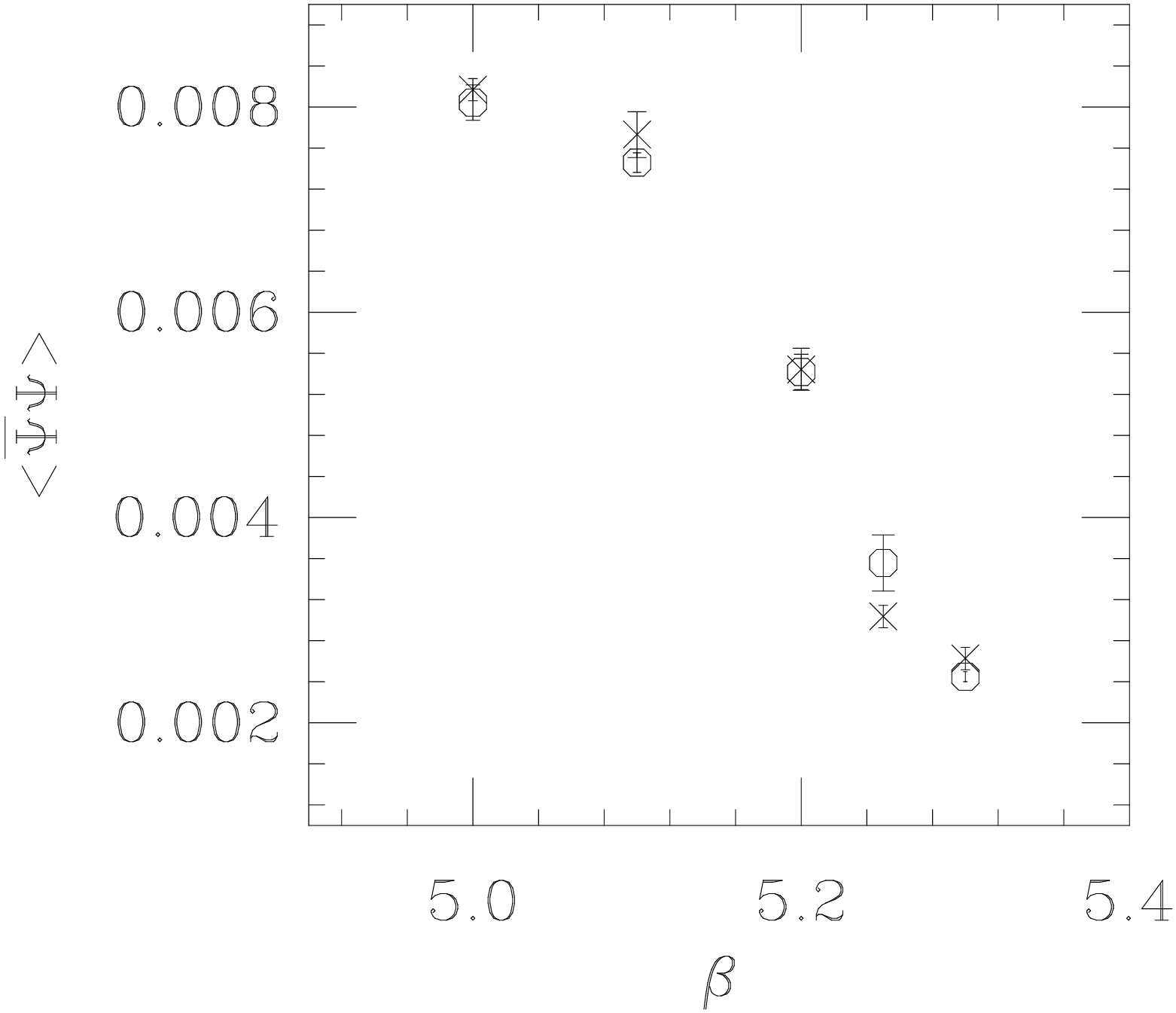}{\ten}{\figsizeA}

A very important question relates to the order of the finite
temperature transition of QCD in nature.  It is known that both Wilson
and staggered fermions with three flavors produce a first order
transition at small quark masses. However, they disagree on the values
of the quark masses where this happens.  Wilson fermions
\cite{trans_order_wilson} predict a second order transition at
physical quark masses while staggered fermions suggest a second order
transition for the physical value of $m_s$ and $m_d=m_u=0$ and a
cross-over for physical values of all three masses
\cite{trans_order_stag}.

Because of the improved chiral properties of DWF it is hoped that
significant progress can be made.  The results of a preliminary study
with three degenerate flavors with $L_s=32$ and $m_f=0.02$ are shown
in figure 10. There is no evidence of a first order transition for
these parameters. A simulation with $L_s=64$ and $m_f=0.01$ will begin
shortly. In tandem, a similar calculation using staggered fermions is
now in progress in order to be able to directly compare the DWF
results at some physical scale with those from staggered fermions at
the same scale.

\section{DWF improvements from the pure gauge sector}

In this section improvements to the $L_s$ behavior of DWF are
discussed. The improvements are attempted by modifying the pure gauge
part of the action. This work is in progress \cite{CU_group}.  For
other promising ideas see also
\cite{Neuberger_Dubna,Shamir_Dubna,Neuberger_bounds}.

As discussed in section 2, there appear to be two general sources that
can slow down the approach to the chiral limit as $L_s$ is
increased. The first is due to fluctuations and the second is due to
configurations in the vicinity of changing topology.  As the continuum
limit is approached both of those sources are suppressed. The first
one since the continuum--like fields are ``smoother'' and the second
one because different topological sectors are separated by
increasingly large energy barriers that suppress topology changes.
The problem is that thermodynamic calculations are necessarily
restricted to relatively strong couplings since the available
computing resources allow access to lattices with only a few lattice
sites $N_t$ along the time direction.  For 2 flavor QCD with DWF and
Wilson plaquette pure gauge action, the transition for $N_t=4$ is at
$\beta=5.325$. As was seen in section 8, $L_s=24$ is not enough to
reduce the chiral breaking effects due to the regulator and as a
result the pion mass is almost three times its physical value. It has
been estimated that in order to bring the pion mass to physical
values, $L_s=64 - 96$ may be needed
\cite{lat98_Fleming,lat99_Fleming}.  Although this is possible with
the 400 Gflops machine at Columbia University
\cite{lat98_NHC,lat99_NHC}, an improvement that will allow the same
degree of chiral symmetry restoration for smaller $L_s$ is very
desirable. It is in this sense that improvements of DWF are discussed.

One could hope to improve DWF by improving the pure gauge part of the
action so that the gauge field configurations will be ``smoother'' and
will contain fewer objects that tend to change topology.  The
connection of DWF with the topology of the gauge field is pronounced
and approximate forms of the index theorem on the lattice have been
found \cite{NV} \cite{EHN_flow}. One could try to
exploit this by manipulating the pure gauge sector and have a good
chance that an improvement in the fermionic sector is achieved.  This
discussion is clearly heuristic and can only serve as a guide to the
choice of actions. Once a choice has been made the effects can only be
determined by explicit numerical simulations.  In this spirit several
actions have been examined.

\subsection{Iwasaki action for quenched QCD at $\beta=5.7$}

The Iwasaki pure gauge action is very appealing since it is expected
to be closer to the continuum action.  A quenched numerical simulation
at $\beta=5.7$ produced the $m_\pi^2$ values shown in figure 11.  For
comparison, the values with a Wilson plaquette action are also shown.
Both sets are the extrapolated $m_f = 0$ values.  The Iwasaki points 
at $L_s=16$ are lower than the Wilson
points at $L_s=48$! This indicates that at these lattice spacings
($a^{-1} \approx 1$ GeV) an important improvement has been found.

\figure{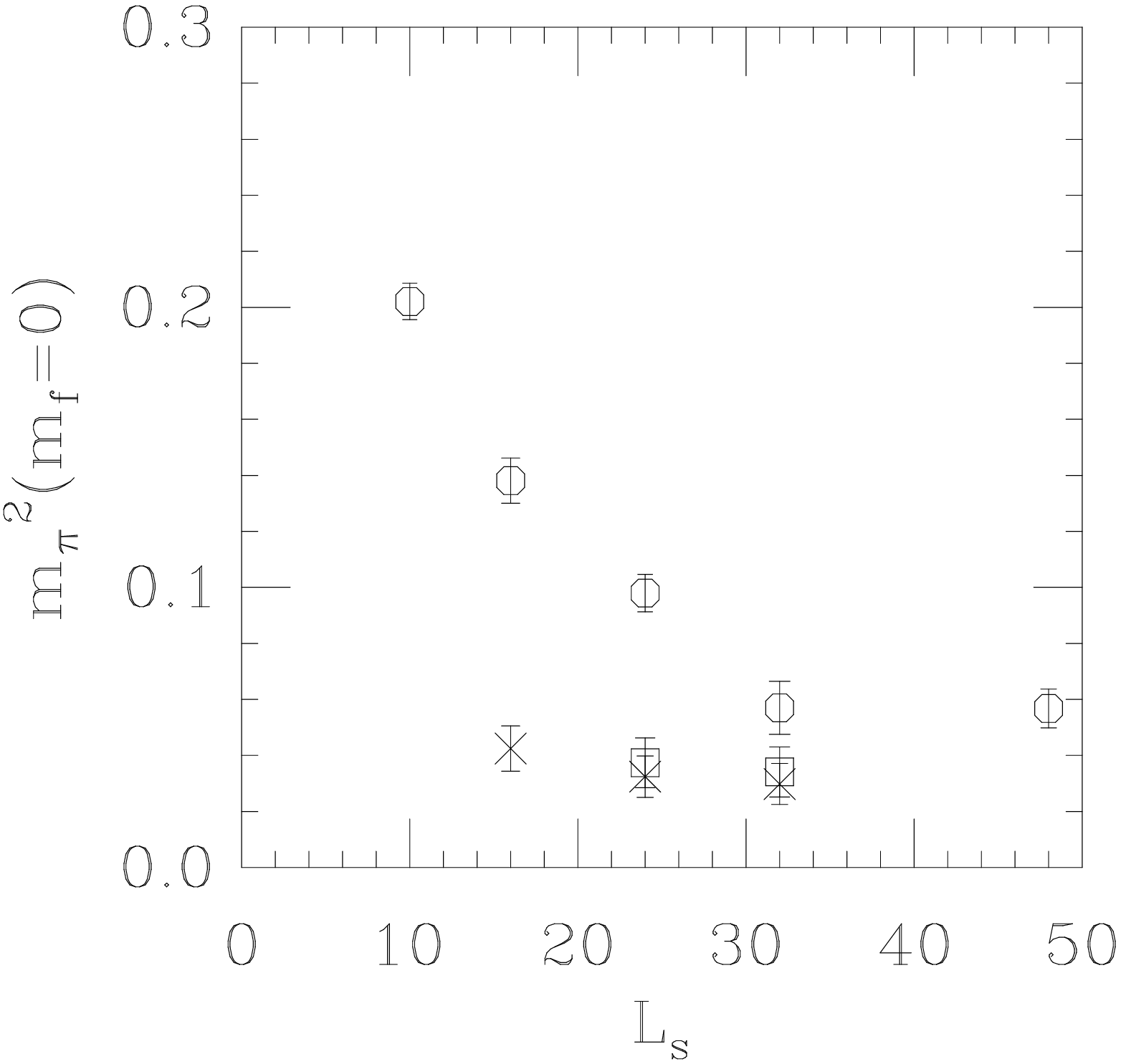}{\eleven}{\figsizeA}

\subsection{Wilson plaquette action}

In what follows the Wilson single plaquette action with dynamical 2
flavor DWF is used as a ``benchmark''. Since dynamical
simulations are very demanding small lattices with size $8^3 \times 4$
are used at $L_s=24$, $m_f=0.02$, $m_0=1.9$ in order to locate the
transition. Then a zero temperature simulation is done on an $8^4$
lattice at the critical coupling in order to measure $\mres$. This
measurement is possible due to the work in 
\cite{lat98_Fleming,lat99_Fleming}.  Here
the more naive $\mres = \PsibarPsi / \chi_\pi - m_f$ is
used. Since we are looking for improvements by a factor of two or
larger this naive estimation is adequate.  The transition is at 
$\beta_c = 5.325$, where $\mres= 0.028(2)$ and $a^{-1} \approx
700$ MeV.

\subsection{Iwasaki action}

From the success at $a^{-1} \approx 1$ GeV of the quenched theory one
could reasonably expect that a similar effect will be present in
dynamical QCD near the transition.  Unfortunately, as already observed
in section 8, this is not the case since the pion mass is about the
same. Also, a measurement of $\mres$ at the transition ($\beta_c=1.9$)
produced $\mres=0.030(3)$ which is similar to the Wilson case. This
failure to improve could be due to the larger lattice spacing at
$\beta_c$ ($a^{-1} \approx 700$ MeV) as compared to $\beta=5.7$
quenched.

\subsection{The restricted plaquette action}

It has been found that small size objects are responsible for most of
the ``unphysical'' topology changes \cite{EHN_flow}. 
%
%
One would expect such objects to be associated with
larger than normal plaquette values. Therefore, one could restrict the
value of $1 - Tr[U_p]/3$ to be less than a certain value and hope to
reduce the presence of such objects. A histogram of $1 - Tr[U_p]/3$
for the Wilson action with dynamical DWF just below the transition is
shown in figure 12a. 
An action with a term 
$( [1 - Tr[U_p]/3] / c)^{20}$
suppresses values of $1 - Tr[U_p]/3$ above the cutoff ``c''. 
For $c=0.8$ it was found
that $\beta_c \approx 1.4$ and $\mres=0.027(1)$.  For $c=0.75$,
$\beta_c$ was less than 0.01 and it did not make sense to pursue
further. Obviously there is no improvement. However, the dependence of
$\beta_c$ on ``c'' is surprising since $c=0.75$ is still in the
tail of the distribution in figure 12a. The distribution for $c=0.8$
is in fig. 12b.

{\epsfxsize=\figsizeB truein
\centerline{\epsffile{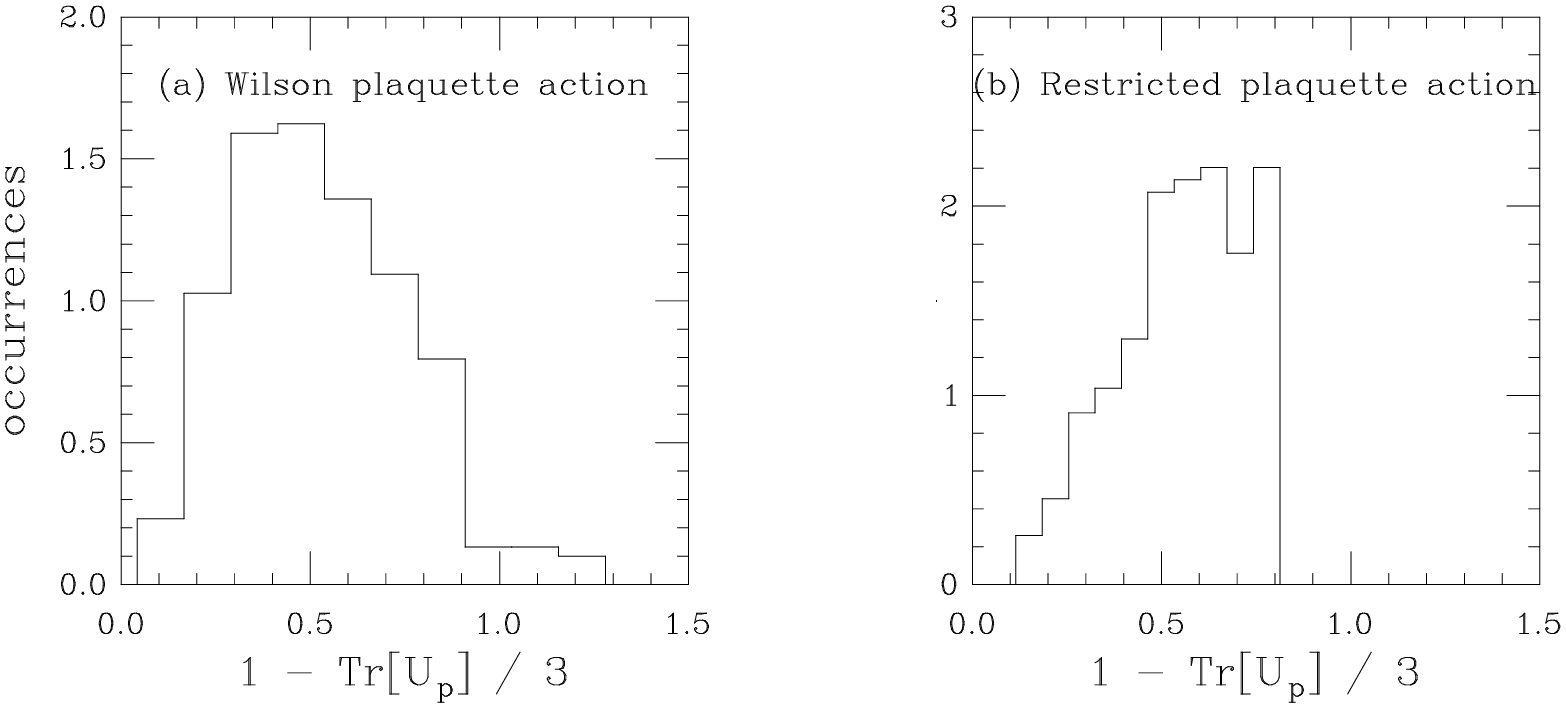}}
\vskip -1.5 truein 
\centerline{\vbox{{\bf \noindent Figure 12.} \twelve}}
\medskip}

\subsection{The enhanced rectangle action}


At large $N_t$ one could imagine that
the transition is driven by instanton type objects of many sizes. Then
removing the objects of size one and less would not affect the transition and
may improve DWF. But at $N_t=4$ one could imagine that only objects of
size one and two exist. Distinguishing them could be very hard. To test 
these ideas the following action that enhances
objects with size $\approx 2$ through the transition by a fixed
amount was used:
\begin{equation}
\beta {Tr[U_p] \over 3} - x {Tr[U_r] \over 3} \ \ ,
\label{xr}
\end{equation}
where $U_r$ is the product of links around the $1 \times 2$
rectangles. This action has the correct continuum limit since the
second term is of order $g^2$. 
The difference from the Iwasaki action is that the
rectangle has a $\beta$ independent coefficient $x=3.0$.  
The transition is at $\beta \approx 15$. At that coupling
$\mres=0.0050(4)$! This large improvement lends support to the above
speculations.

Furthermore, it provides us with another picture
\cite{thanks_vanBall_Kovacs_Stamatescu}. As is known for the
Iwasaki action if a large instanton begins to shrink it will encounter
an energy barrier and will not be able to shrink below one to two
lattice spacings. These ideas were developed in relation to the
cooling methods but they are clearly relevant here. These energy
barriers may make it more difficult for instanton-like objects to
shrink below the lattice spacing and therefore change topology. This
could be an alternative explanation for the improvement seen with
Iwasaki action at quenched $\beta=5.7$ ($a^{-1} \approx 1$ GeV). Also,
since it is known that the more negative the coefficient of the
rectangle the larger the barrier, the observed improvement with the
action of eq. \ref{xr} at $a^{-1} \approx 700$ MeV is not unexpected.
Currently, an Iwasaki type action with $c=-1.0$ as in
\cite{Iwasaki_Yoshie} is being investigated \cite{CU_group}.

\section{DWF improvements from the fermion sector}

Other improvements can be attempted by changes in the
fermion sector.

As discussed earlier, the topology changing configurations are the ones
for which the transfer matrix has a unit eigenvalue.  It is known that
this is equivalent with the traditional Wilson Dirac operator with
mass equal to $-m_0$ having a zero eigenvalue 
\cite{NN1,Furman_Shamir}. Therefore, including extra species of
Wilson fermions with Dirac operator
$D^\dagger D + h^2$ may reduce these eigenvalues depending
on the value of the parameter $h$. This work is in progress \cite{CU_group}.

\section*{Acknowledgments}

This research was done on the Columbia QCDSP 
\cite{lat98_NHC,lat99_NHC} and was supported
by DOE grant 
\# DE-FG02-92ER40699 and NSF grant \# NSF-PHY96-05199.
I am grateful for discussions on DWF improvements
with:
P. van Baal,
T. Kovacs,
H. Neuberger,
Y. Shamir,
I. Stamatescu,
and E.T. Tomboulis.


\vfill

\begin{thebibliography}{99}

\bibitem{Kaplan0} D.B. Kaplan, 
Phys. Lett. B 288 (1992) 342.

\bibitem{Kaplan1} D.B. Kaplan, 
Nucl. Phys. B (Proc. Suppl.) {\bf 30} (1993) 597.

\bibitem{DWF_reviews} 
R. Narayanan, Nucl. Phys. {\bf B34} (Proc. Suppl.)  (1994) 95; 
M. Creutz, Nucl. Phys. {\bf B42} (Proc. Suppl.)  (1995) 56; 
Y. Shamir, Nucl. Phys. {\bf B47} (Proc. Suppl.)  (1996) 212; 
T. Blum, Nucl. Phys. {\bf B73} (Proc. Suppl.)  (1999) 167; 
H. Neuberger, Lattice 99 proceedings, hep-lat/9909042;
M. L\"{u}scher, Lattice 99 proceedings, hep-lat/9909150;

\bibitem{Neuberger_Dubna} H. Neuberger, these proceedings, hep-lat/9912013.

\bibitem{Shamir_Dubna} Y. Shamir, these proceedings, hep-lat/9912027.

\bibitem{DWF_Dubna} These proceedings:
W. Bietenholz, hep-lat/0001001; 
A. Borici, hep-lat/9912040;
M. Creutz, hep-lat/9912006; 
U. Heller, hep-lat/9912043, hep-lat/9912042;
I. Horvath, hep-lat/9912030;
W. Kerler; 
A.A. Slavnov; 
M. Wingate.

\bibitem{Shamir} Y. Shamir, 
Nucl. Phys. B {\bf 406} (1993) 90.

\bibitem{Furman_Shamir} V. Furman, Y. Shamir, 
Nucl. Phys. {\bf B439} (1995) 54.

\bibitem{PMV} P.M. Vranas, 
Lattice 96, Nucl. Phys. {\bf B53} (Proc. Suppl.) (1997) 278;
Phys. Rev. {\bf D57} (1998) 1415.

\bibitem{NN0} R. Narayanan, H. Neuberger, 
Phys. Lett. B {\bf 302} (1993) 62.

\bibitem{NN1} R. Narayanan, H. Neuberger, 
Phys. Rev. Lett. {\bf 71} (1993) 3251;
Nucl. Phys. B {\bf 412} (1994) 574; 
Nucl. Phys. B {\bf 443} (1995) 305.

\bibitem{EHN_flow} 
U.M. Heller, R. Edwards and R. Narayanan, 
Nucl. Phys. {\bf B535} (1998) 403;
Phys.Rev. D60 (1999) 034502.

\bibitem{NNV} R. Narayanan, H. Neuberger and P. Vranas, 
Phys. Lett. B {\bf 353} (1995) 507;
Nucl. Phys. B (Proc. Suppl.) {\bf 47} (1996) 596.

\bibitem{VKT} P.M. Vranas, I. Tziligakis and J. Kogut,
to appear in Phys. Rev. {\bf D}, hep-lat/9905018.

\bibitem{Aoki_phase} S. Aoki, Phys. Rev. {\bf D30} (1984) 2653;
S. Aoki, S. Boettcher and A. Gocksch,
Phys. Lett. {\bf B331} (1994) 157.

\bibitem{Aoki_phase_DWF} S. Aoki,
Lattice 99 proceedings, hep-lat/9909154.

\bibitem{CU_zero_modes} P. Chen et.al., Phys. Rev. {\bf D59} (1999) 054508.

\bibitem{lat98_Fleming} G. Fleming et.al., 
Nucl. Phys. {\bf B73} (Proc. Suppl.) (1999) 207.

\bibitem{lat98_Kaehler} A. Kaehler et.al., 
Nucl. Phys. {\bf B73} (Proc. Suppl.) (1999) 405.

\bibitem{lat99_Fleming} G. Fleming, 
Lattice 99 proceedings, hep-lat/9909140.

\bibitem{EHKN_quenched_thermo} 
R. Edwards, U. Heller, J. Kiskis, and R. Narayanan, hep-lat/9910041.

\bibitem{lat99_Sinc} J.F. Lagae and D.K. Sinclair, 
Lattice 99 proceedings, hep-lat/9909097.

\bibitem{lat98_Vranas} P.M. Vranas et.al., 
Nucl. Phys. {\bf B73} (Proc. Suppl.) (1999) 456.

\bibitem{ICHEP_NHC} N. Christ et.al., 
contribution to ICHEP 98, hep-lat/9812011.

\bibitem{dpf99_Vranas} P.M. Vranas, 
contribution to DPF 99 proc., UCLA, hep-lat/9903024.

\bibitem{lat99_Vranas} P.M. Vranas, 
Lattice 99 proceedings, hep-lat/9911002.

\bibitem{PANIC99_Fleming} G. Fleming, 
contribution to PANIC 99, Uppsala, Sweden, hep-ph/9910453.

\bibitem{CU_omega} S. Chandrasekharan, D. Chen, N.H. Christ, W. Lee,
R. Mawhinney, and P.M. Vranas, 
Phys. Rev. Lett. {\bf 82} (1999) 2463.

\bibitem{Kogut_Lagae_Sinc} J.B. Kogut, J.F. Lagae, D. K. Sinclair, hep-lat/9801020.

\bibitem{Pizarski} R. Pizarski and F. Wilczek, 
Phys. Rev. {\bf D29} (1984) 338.

\bibitem{lat99_Wu} L. Wu, 
Lattice 99 proceedings, hep-lat/9909117.

\bibitem{lat99_RDM} R. Mawhinney, 
Lattice 99 proceedings.

\bibitem{lat99_Karsch} F. Karsch,
Lattice 99 proceedings, hep-lat/9909006.

\bibitem{Iwasaki} Y. Iwasaki, 
Nucl. Phys. B {\bf 258} (1985) 141. 

\bibitem{CU_group} In progress,
P. Chen,
N. Christ,
C. Cristian,
G. Fleming,
A. Kaehler,
T. Klassen,
X. Liao,
G. Liu,
C. Malureanu,
R. Mawhinney,
G. Siegert,
C. Sui,
L. Wu,
and
Y. Zhestkov.

\bibitem{trans_order_wilson} Y. Iwasaki et.al.,
Phys. Rev. {\bf D54} (1996) 7010.

\bibitem{trans_order_stag} F. Brown et.al, 
Phys. Rev. Lett. {\bf 65} (1990) 2491.

\bibitem{Neuberger_bounds} H. Neuberger, hep-lat/9911004.

\bibitem{lat98_NHC} N. Christ, et.al.
Nucl. Phys. {\bf B73} (Proc. Suppl.) (1998).

\bibitem{lat99_NHC} N. Christ,
Lattice 99 proceedings, hep-lat/9912009.

\bibitem{NV} R. Narayanan, P. Vranas, 
Nucl. Phys. {\bf B506} (1997) 373.

\bibitem{thanks_vanBall_Kovacs_Stamatescu}
I am grateful to P. van Baal, T. Kovacs and I. Stamatescu
for pointing this out to me.

\bibitem{Iwasaki_Yoshie} Y. Iwasaki and T. Yoshi\`{e},
Phys. Lett. {\bf B131} (1983) 159.

\end{thebibliography}
\end{document}